\documentclass[aip,amsmath,amssymb,reprint,superscriptaddress]{revtex4-1}
\usepackage{graphicx}
\usepackage{amsmath}
\usepackage{dcolumn}
\usepackage{bm}
\usepackage{bbold}
\usepackage{color}
\usepackage{caption}
\usepackage{subcaption}
\usepackage{mathrsfs}

\begin{document}

\title{Advances in Development of {Quartz Crystal Oscillators at Liquid Helium Temperatures}}

\author{Maxim Goryachev}
\email{maxim.goryachev@uwa.edu.au}
\affiliation{ARC Centre of Excellence for Engineered Quantum Systems, University of Western Australia, 35 Stirling Highway, Crawley WA 6009, Australia}

\author{Serge Galliou, Jo\"{e}l Imbaud, Philippe Abb\'e}
\affiliation{Department of Time and Frequency, FEMTO-ST Institute, ENSMM, 26 Chemin de l'\'{E}pitaphe, 25000, Besan\c{c}on, France}


\begin{abstract} This work presents some recent results in the field of liquid helium { bulk acoustic wave} oscillators. The discussion covers the whole development procedure starting from component selection and characterization and concluding with actual phase noise measurements.
The associated problems and limitations are discussed. The unique features of obtained phase noise power spectral densities are explained with a proposed extension of the Leeson effect.
\end{abstract}

\maketitle

\section{Introduction}

Nowadays, the best short-term frequency stability performance (typically for averaging times less than $1$ minute) { among room temperature solutions available on the market} is achieved with quartz crystal oscillators. { State-of-the-art quartz crystal oscillators can exhibit frequency stabilities lower than  $\sigma_{y}(\tau) = 6\times10^{-14}$ at $5$ MHz, in terms of Allan deviation of the fractional frequency $y$, over integration times $1s < \tau < 100s$\cite{Kuna}. Their short term stability is limited by the electronic noise floor whereas their long term stability is mainly degraded by drift and ageing. Between both areas, stability performance is limited by the frequency flicker noise level of the oscillator loop.} But now it is generally believed that this family of frequency sources operates almost in saturation and reasonable efforts in the future could not lead to considerable stability improvements. So, a jump in technology is necessary. The possibility of replacement of quartz oscillators with MEMS and chip-scale atomic clocks is under discussion\cite{QvsCSAC}. As for long term stability performance, many different types of frequency sources compete including optical and microwave atomic clocks. { For these devices, atomic transitions are used to achieve outstanding long term stability and accuracy. They are usually combined with a local oscillator, that is or can be stabilized to improve the short term stability of the overall system. Atomic transitions in the optical domain lead to the optical clock family which exhibits outstanding frequency stabilities benefiting from their high operational frequencies. Their stabilities, dominated by white frequency noise, are in the order of  $\sigma_{y}(\tau) = 1\times10^{-14} \tau^{-1/2}$ over a very wide range of $\tau$, from less than $1$s up to more than $10^4$s. It can be better according to the ion or neutral atom type: $2.8\times10^{-15} \tau^{-1/2}$ with Al$^{+}$ \cite{Chou1,Chou2}. But optical frequency standards are still complex research-oriented systems only developed in the major laboratories of the domain (NIST, PTB, NPL, SYRTE, NRC, etc), and cannot easily be transported. Apart from these systems, which are still in an exploration status, atomic transitions in the microwave domain have been intensively studied and developed to give robust frequency standards such as cesium (Cs) clocks and hydrogen masers. State-of-the-art, laser-cooled, Cs-fountain microwave clocks have demonstrated impressive frequency measurement accuracy, with fractional uncertainties below the $10^{-15}$ level \cite{bize,weyers,guena}. Commercial Cs-clocks exhibit long term stability of $10^{-14}$ over $10$ days and more. The $\sigma_{y}(\tau)$ plot of commercial hydrogen masers typically shows a minimum of a few $10^{-15}$ at $10^3$s.} 

{Among this variety of frequency standards based on atomic transitions, the microwave} cryogenic sapphire oscillators { (CSO) is out of this category but} can all be found in high long-term stability regimes. {As an example, CSOs  demonstrated long term frequency drift less than  $5\:10^{-14}/$day coupled with exceptional short term stabilities in the order of  $\sigma_{y}(\tau) = 1\;10^{-15} \tau^{-1/2}$ over $1 < \tau < 10s$ \cite{Nitin}, even achieved with transportable and commercial units\cite{Grop}.
Like dielectric resonators used in CSOs, acoustic resonators (especially quartz bulk acoustic wave resonators) can potentially lead to significant technology improvement despite their quite low frequencies.}

Usually, quartz oscillator performance over long integration times is very poor, { at least at room temperature}. However, cryogenic quartz systems (or more generally, systems based on liquid helium bulk acoustic wave devices) remain a dark horse in the race for the best stability in the short, middle, and moderately long terms.  { Although extremely high quality factors of quartz devices at cryogenic temperatures have been demonstrated\cite{quartzPRL,Cryo2S, ApPhLt, apl2012},} until now cryogenic quartz systems have been rather under-investigated, and their potential and perspectives are not generally known. {Some existing studies in the field of work dedicated to these devices and systems have not been systematic, and most studies are now outdated, having been based on outdated cryogenic technologies.} This work presents some advances and systematic research results in the field of liquid helium oscillators { based on Bulk Acoustic Wave (BAW) resonators in quartz crystal}.

{ Investigation of liquid helium quartz crystal oscillators is interesting} {for several reasons. First, intrinsic mechanical losses of quartz dramatically decrease below $10$K, as already mentioned, and  becomes comparable or lower than dielectric losses of cryogenic sapphire, even if their respective operating frequencies $f_{0}$ are not in the same range:  quartz crystal resonators can exhibit unloaded quality factor higher than 1 billion at $f_{0} = 200$ MHz \cite{ScRep}. Second, quartz resonator flicker noise is related to mechanical losses: it has been demonstrated, at least from experiments, that the Power Spectral Density (PSD) level at $1$~Hz, $S_{y}(f \text{=1Hz})$, of the fractional frequency $y = df_{0}/f_{0}$, is proportional to $1/Q^{n}$ with $n\approx4$ \cite{Gagnepain, Parker, Planat}. In addition, investigation on this topic} helps to understand whether or not the BAW technology at ambient temperatures has reached its limits. { Quartz material is the best candidate in this field, quartz resonators being the more mature BAW-type resonators. Indeed, it is still relevant} {to predict the frequency stability limits of quartz-based oscillators in comparison with other technologies. }

This article is composed as follows. Section \ref{cryoo} introduces the reader to the cryogenic system used, and the cryogenic environment created. Next, Section~\ref{CS} presents results on the investigation of different components for liquid helium oscillators. Their optimal choice is explained. The next section gives implementation details and measurement results for oscillator design based on a field-effect transistor. Section~\ref{addii} presents a generalisation of the Leeson model that explains the measured spectra. Section~\ref{HBT} describes the design based on a heterojunction bipolar transistor. Section~\ref{SLDD} discusses the main limitations of the frequency stability of liquid helium cooled BAW oscillators. Final remarks are made in the conclusion.

\section{Cryogenic Environment}
\label{cryoo}

\begin{figure}[!h]
	\centering
	\includegraphics[width=3.25in]{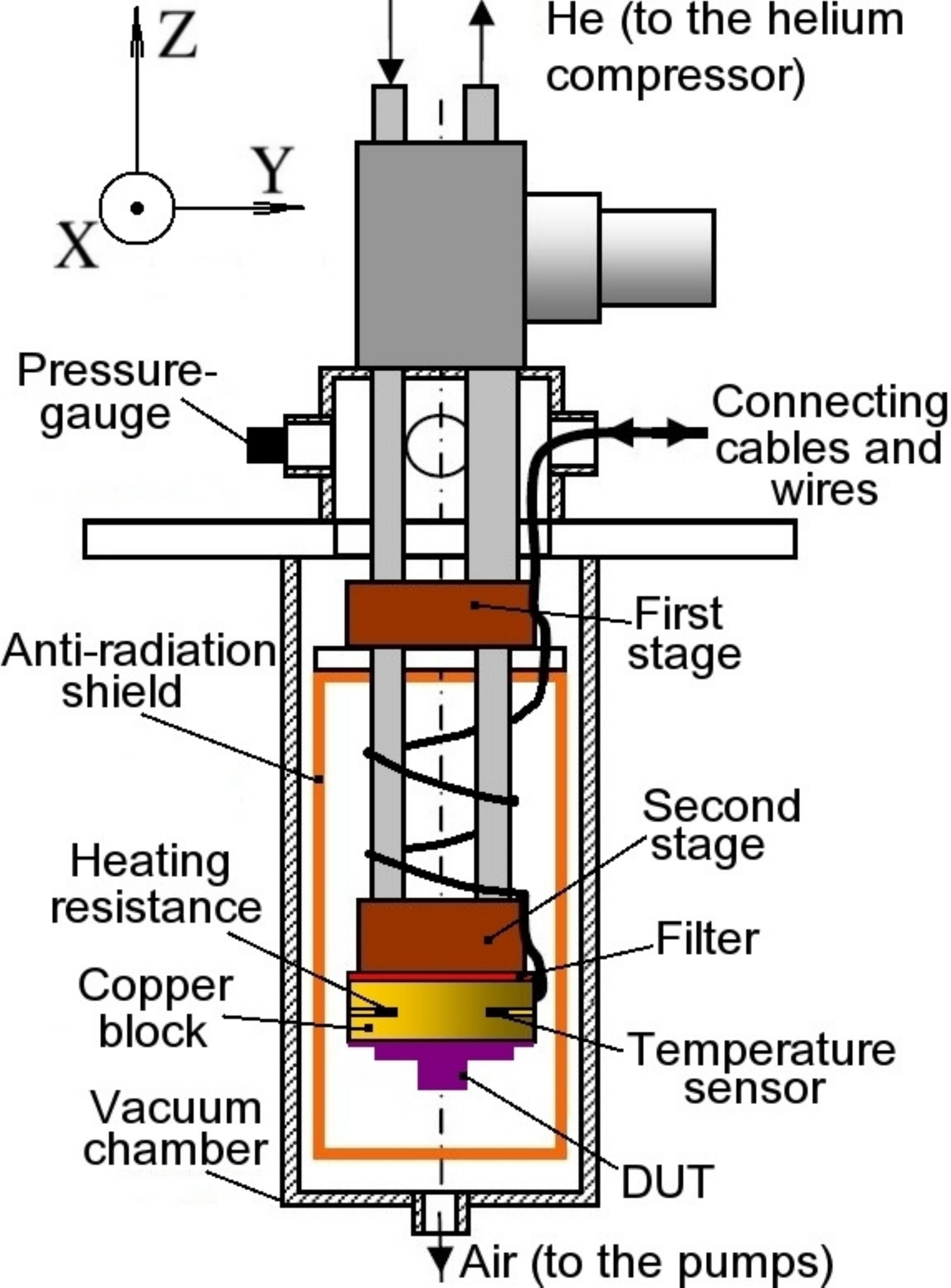}
	\caption{\label{P001FZ}Cryogenic system}
\end{figure}

In this work, a two-stage pulse-tube cryocooler (see Fig.~\ref{P001FZ}) is used as the cryogenic system core. The cryocooler is shielded with two protective screens: a vacuum chamber and an anti-radiation shield. The former prevents temperature losses due to air heat conduction. The air pressure inside the chamber is about $10^{-7}$~mBar. This shield is at room temperature and is made with a light-reflecting material. Both primary and secondary pumps are cut off from the chamber during the whole measurement time. The anti-radiation shield is attached to the first cryocooler stage and its temperature is about $50$K. Its inner and outer surfaces are covered with a specially fabricated light-reflecting paper with a series of mylar sheets to reduce radiation losses of the second stage.

Nominally, the cryocooler absorbs up to $1$~W at its second stage at $4$K. The second stage temperature can go down to almost $3$K in the unloaded configuration. The thermal losses are mainly introduced with connecting cables needed for measurements and environmental control. The {device under test} (DUT) is attached to a temperature-controlled copper block. 




 The almost periodic temperature fluctuations of the cryocooler second stage due to the helium pulsation are about $0.3$K. The power spectral density (PSD) of temperature fluctuations of the second stage in the steady state consists of an $f^{-2}$ region for frequencies less than about $10^{-2}$~Hz and an $f_0$ region for higher frequencies. In addition, a clear spurious frequency of $1.7$~Hz and its harmonics are present due to temperature pulsations initiated by the pulse tube cycle.

In order to improve the thermal environment of the DUT, passive and active temperature stabilization methods are used. First, a thermal filter is introduced between the second stage and the copper block (see Fig.~\ref{P001FZ}). The filter reduces the temperature pulsation down to $15$~mK, but increases the absolute temperature difference between both parts. This difference mainly depends on the number of cables connected to the DUT. Second, active temperature regulation is implemented with a Lakeshore Model 332 Temperature
Controller, a CERNOX temperature sensor, and a $20$~Ohm heating resistor. The sensor and the heater are installed inside the copper block under the filter. The control system eliminates the $f^{-2}$ region in the temperature fluctuation spectrum (at least for frequencies higher than $10^{-4}$~Hz) and reduces the temperature fluctuation down to $5$mK. The power spectral densities of temperature fluctuations in both the second stage (unregulated and non-filtrated part) and the DUT (after filtering with active regulation) can be found in\cite{PhNoCr}.

The power spectral density of temperature fluctuations shows that the remaining noise due to temperature instabilities at the DUT level (with no thermal load or cables installed) is almost a white noise in the frequency range of $10^{-3}-10^0$~Hz. It should be noted however that the $f^{-2}$ spectrum slope can still appear at this range of frequencies at the DUT level when measurements are performed during exponential transients in the whole thermo-mechanical system (cryogenerator, cables, DUT, etc.). This happens when the thermal load is high enough, and connection cables create a thermal link between regulated and unregulated parts of the system.

Another type of cryocooler environmental disturbance is vibration. According to the cryocooler specification, the permanent pulsation of helium inside the pulse-tube causes a DUT vibration with an amplitude of about $10~\mu$m and a frequency $f=1.7$~Hz. The spectrum of vertical DUT acceleration (Z-axis in Fig.~\ref{P001FZ}) has been calculated, based on measurements made with a high sensitivity piezoelectric accelerometer (Model A/600) and a high speed real-time ADwin-Pro II\cite{ADwin} system with a custom-made low noise charge amplifier. The DUT acceleration PSD along the vertical axis is shown in\cite{PhNoCr}. In fact, the sensitivity of the accelerometer at cryogenic temperature is unknown. So, the presented data cannot be graduated.


Nevertheless, the measurement suggests that the DUT is subject to $f^{-1}$ instabilities in the frequency range of $10^{-2}-10^1$~Hz due to external vibration in the vertical axis. It has to be noted that no vibration compensation system is installed for the presented measurements. 

\section{Component Selection}
\label{CS}

One of the most challenging tasks for designing electronic systems for liquid helium temperatures is a choice of passive and active components. Indeed, the majority of the available electronic components have a reliable temperature range from $-40$ to $85^\circ$C (industrial grade) or from $-55$ to $125^\circ$C (military grade) in the best case. Moreover, the majority of work dedicated to cryogenic electronics deal with the liquid nitrogen range (about $77$K)~\cite{Low}. At the same time, the liquid helium temperature range (about $4.2$K) is much less investigated. So, the procedure of electronic design has to have the following steps: 1) finding the components which are able to work at the given frequencies, 2) their characterization, and 3) appropriate component choice. This section gives some results obtained by the authors in this field. 

\subsection{BAW Resonators}

It has been known for a long time that BAW resonators exhibit high quality factors at low temperatures\cite{Old1,Old2,Old3, Smagin}. Investigations in this field were continued by obtaining even higher $Q$-values\cite{Mossuz, Suter, Phnoise2, Habiti, Habiti2}. Recently, systematic research of different properties of BAW devices at liquid helium temperatures has commenced at FEMTO-ST Institute, Besan\c{c}on, France. As a part of the research, the different operational mode are studied and compared, paying attention to frequency-temperature characteristics\cite{Imbaud, Cryo2S}. Based on measurement data, the loss phenomena in quartz resonators have been analyzed\cite{ApPhLt}. Furthermore, amplitude-frequency nonlinear effects, temperature sensitivity, LGT resonator investigation and mode comparison~\cite{SunFr} have been also subjects of this study. In addition to this, the resonator phase noise at the given condition has been measured\cite{PhNoCr}.

The choice of the resonator was fixed upon the BVA (Bo\^{\i}tier \`{a} Vieillissement Am\'{e}lior\'{e}) resonators\cite{reson2}, because it has been already shown that electrodeless resonators exhibit higher quality factors than electrode-deposited resonators\cite{Suter}. More precisely, SC-cut resonators\cite{reson1} optimized to work on the 3$^{\text{rd}}$ overtone of the C-mode ($5$~MHz) at $80^\circ$C (at this temperature, the frequency-temperature characteristic of such a device exhibits a turn-over point) are chosen. These devices obtain the highest measured quality factors amongst all BAW devices\cite{SunFr}. 
 
\subsection{Passive Components}

Two types of $1\%$-accuracy resistors available at the laboratory were tested at cryogenic temperatures:\begin{enumerate}
\item {\it thick film process resistors} from YAGEO (PHYCOMP) - RC0805JR-SKE24L (SMD Resistors 0805, $1$\%, $150$~V, $100$~ppm/C, power dissipation $0.125$~W).
 \item {\it metal film resistors} from Basic Kit LMA-24-SBR02 (MINI-MELF 0204-50, $1$\%, $50$~ppm/K, rated power dissipation $0.25$~W);
 \end{enumerate}

These two types of $1\%$-accuracy chip resistors have been tested with no superconductive properties found. Contrary to that, some of the verified resistor types demonstrated growth of their resistance. The following general rule has been observed: the higher room temperature resistance value, the higher increase of this value at liquid helium temperatures.

Thick film resistors tested up to $10$~Ohm demonstrate no considerable growth (less than $1\%$). Values of resistances higher than $1$~kOhm may be shifted by more than a couple of dozen percent. And resistors of hundreds of kOhm demonstrate enormous increase in the considered parameter (more than $100\%$). This unpleasant property may be overcome by simply taking into account of these changes of resistance value {\it a priori}. But this approach is complicated by the high dispersion of values. So, sufficient accuracy of the devices cannot be achieved. Moreover, the resistance values change significantly in the working temperature range $3.6$-$5$K: tens of percent for resistances of hundreds of kOhm. This fact makes these devices inapplicable in systems where low temperature sensitivity (such as ultra stable reference sources) is important.

Metal film resistors demonstrate the same pattern of resistance change, but with much smaller deviations (for example, less than $2\%$ for resistors of hundreds of kOhm). In the temperature range $3.2$-$5$K the deviation does not exceed $1\%$ for resistors less than $700$~kOhm. So, this type is preferable for sensitive low temperature applications.

Values of all capacitors also change at cryogenic temperatures. Capacitors with two common types of ceramic dielectric materials were tested (from different manufacturers):
\begin{enumerate}
 \item {\it EIA Class 2 dielectric}, in particular X7R, that is designed for capacitors with typically good non-critical coupling, filtering, transient voltage suppression, and timing applications;
\item {\it EIA Class 1 dielectric}, in particular C0G (EIA) or NP0 (industry spec.), that is the material with the lowest capacitance/temperature dependence (Negative-Positive zero). C0G/NP0 dielectrics have the lowest losses.
 \end{enumerate}
 
All samples of X7R capacitors exhibit an almost tenfold decrease in capacitance going from room to liquid helium temperatures. For example, capacitance of the MLT030227 CMS1206 device changes from $3.3$~nF to $268$~pF. Moreover, starting from approximately $500$~pF and lower, capacitors of this type exhibit inductive behavior at the considered cryogenic conditions. 

C0G capacitors demonstrate no considerable change of values between room and liquid helium temperatures as well as within $3.8-5$ temperature range. For example, the capacitors (from VISHAY, CMS0805, $\pm5\%$) with the nominal value of $22$~pF and measured $25$~pF exhibit $24$~pF at $4$~K. For $100$~pF nominal capacitors, these values are $106$~pF and $103$~pF respectively.
And for $150$~pF samples, $156$~pF is measured at room temperature, and $153$~pF is obtained for $4K$.

Comparing these two solutions of dielectric choice, C0G capacitors are preferable since they do not require any value correction or compensation in the course of the system design.

Ordinary electrolytic capacitors, which in the considered applications are usually used for voltage stabilization and do not require extraordinary accuracy, do not show a considerable value shift which has to be corrected.

\subsection{Active Components}

Amongst all possible candidates for the role of an active device at liquid helium temperatures, the following transistor types are chosen: SiGe HBT, N-Channel JFET, and N-Channel Dual Gate MOSFET. The main characteristics as well as the approaches for modelling and measurement of HBTs (in particular, BFP650) are described in~\cite{My1}. The advantages of this device are the high current gain, high cut-off frequency, availability, and compact size. The main disadvantage is their low robustness { to cooling and heating cycles. Different expansion and contraction coefficients of different parts of the device can lead to considerable stress. 
After a few cycles of cooling and heating this led to irreversible changes.}. Nevertheless, the workability of these devices was proved by successful amplifier design for liquid helium temperatures\cite{My1}.

Another possible solution is the N-Channel JFET. The reference tested (2N4223) has amplification capabilities at liquid helium temperature conditions, but only without direct thermal contact with the cryogenerator second stage. Such a property is due to the self-heating effect. Since these devices did not show reliable and repeatable performance, they are discarded from further utilization.

Another type of field effect transistor that could be used at liquid helium temperatures is the MOSFET. The reference tested (BF988) showed high reliability, but relatively low gain at the desired temperatures. Another disadvantage of this device is that it becomes obsolete. The applicability was also proved by a successful amplifier design.

\section{Procedure of Oscillator Design for Liquid Helium Temperatures}

The process of design and implementation of liquid helium BAW oscillator is associated with several issues that do not exist for room temperature procedures.

The first difficulties come at the level of oscillator modelling. High quality factors of quartz resonators imply extremely high simulation times and severe problems with convergence in modelling algorithms. The problem can be solved with the application of { a behavioral model instead of a conventional compact model}\cite{My2}. Moreover, the absence of adequate models for the resonator phase noise makes verification of the results with commercial software difficult. Such a model is proposed in \cite{precis}.

Development of an actual cryogenic oscillator is hindered by many technical problems. First of all, this process is very limited by the impossibility of modifying the oscillator electronics ``on-the-fly" which is absolutely required by the complexity of such systems. This means that once the DUT is installed and closed, and the temperature is decreased, one cannot change anything inside the cryocooler. In order to make any modifications the cryocooler has to be stopped, and its temperature has to be increased up to room temperature. This process takes at least $12$ hours. Once modifications are done, the DUT has to be cooled down again. This
process takes about $3.5-4$ hours. This fact means that any schematic modification and its verification could be made only once in $24$ hours. In particular, it makes it almost impossible to utilise the mode-selective circuit in an oscillator loop, because it requires extremely accurate adjustment on an actual device. So, the development process of a cryogenic electronic system (and oscillators in particular) is a very time-consuming and exhausting process compared to the room temperature situation. Moreover, cryogenic devices are much less reliable, their behaviour is far less predictable, and the environmental influences are much less known. So, the performance of the first cryogenic crystal oscillators can hardly be compared with performance of the state-of-the-art room temperature devices developed over many decades.

It has to be noted that even though the sensor temperature reaches $4$K after $4$ hours of cooling process, frequency stability measurements have to be made after at least additional $20$~hours. This is due to very long-term time constants of the cryogenerator thermal processes.

Another feature of the cryogenic oscillator development is the necessity to design a special thermo-mechanical structure. This requirement is a consequence of the many issues concerning resonator and transistor temperatures, dissipated power, temperature instabilities, and so on.

In addition, crystal resonators working at cryogenic temperatures have many overtones and modes which can be potentially excited. And since, as it is mentioned above, the utilization of a mode-selection circuit is not possible, one cannot predict an oscillation mode \textit{a priori}.

\section{MOSFET-Based Design}
\label{MOSFET}

\begin{figure*}[!ht]
	\begin{center}
	\includegraphics[width=0.69\textwidth]{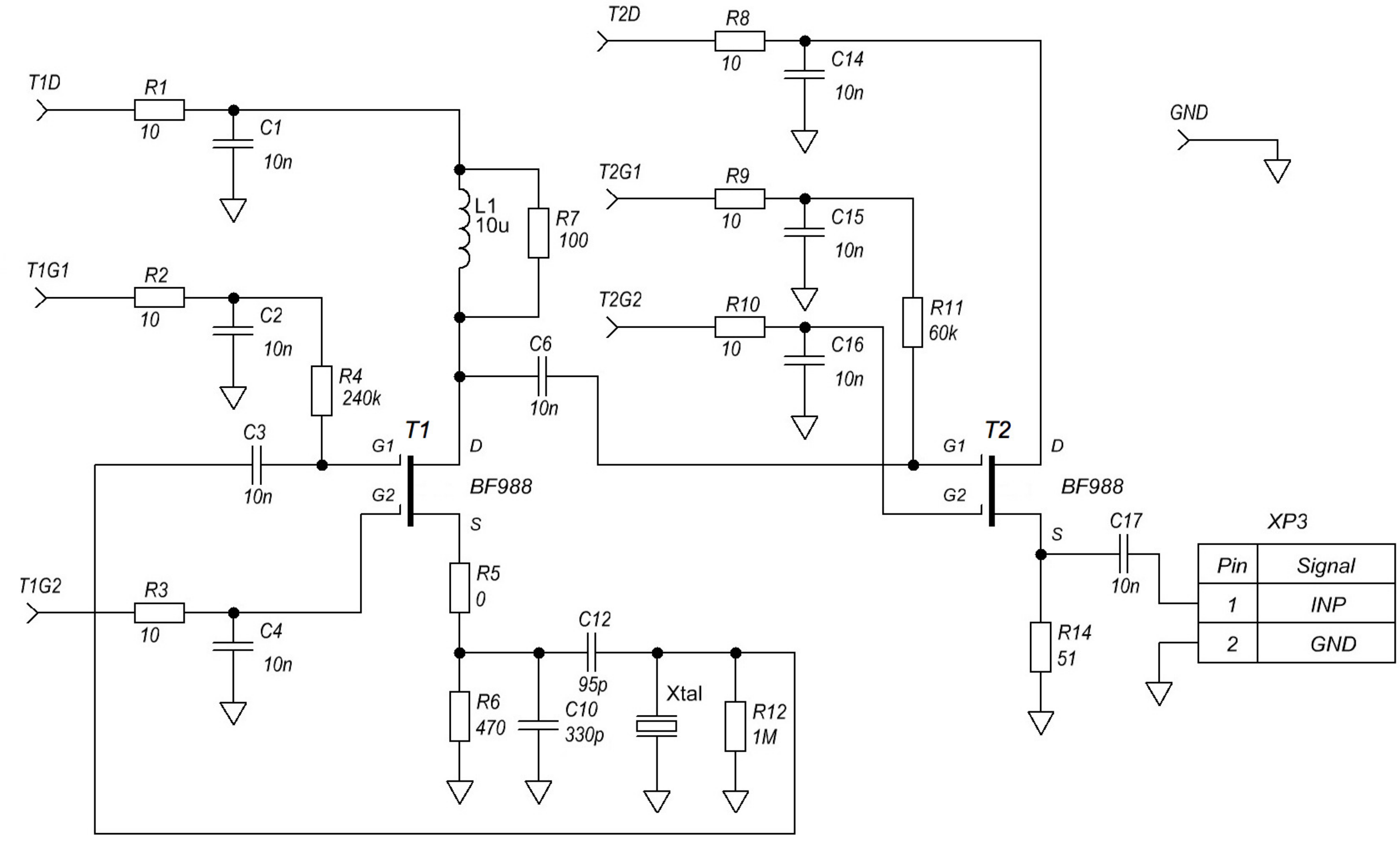}
	\caption{Schematic diagram of a cryogenic crystal oscillator based on a MOSFET transistor.}
\label{D003DO}
	\end{center}
\end{figure*}

The first liquid helium oscillator design was implemented based on robust MOSFETs. The schematic diagram is shown in Fig.~\ref{D003DO}. The oscillator incorporates two transistors. Transistor T1 is a base for a sustaining amplifier inside the oscillating loop. The second transistor {(T2)} is an active element of a buffer that separates the oscillating loop from the connecting cables. These cables could become a primary source of disturbances since they are not temperature regulated.

\begin{figure}[!ht]
	\centering
	\includegraphics[width=0.45\textwidth]{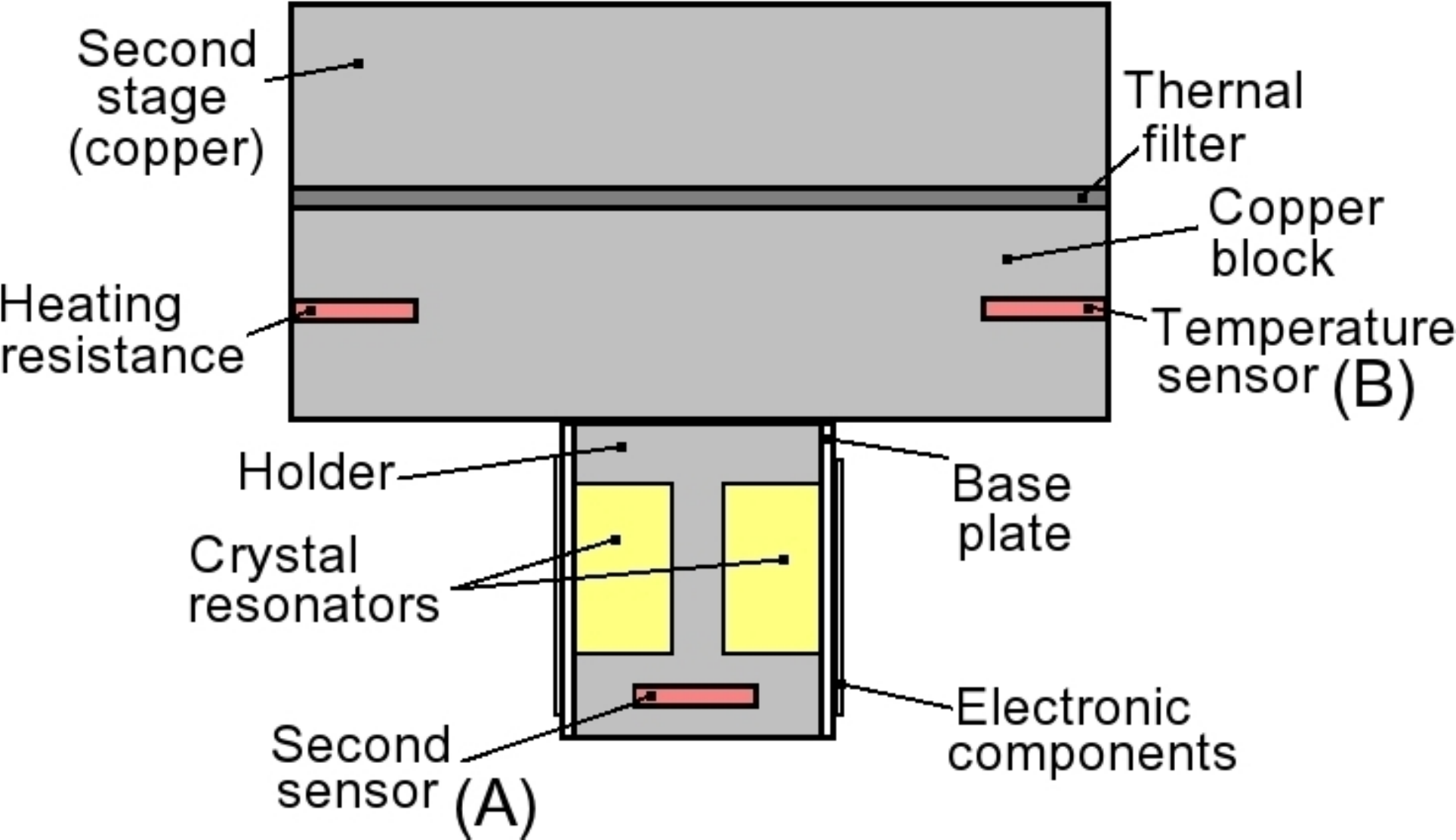}
	\caption{Schematic view (cables and wires are not shown).}
\label{D001DO}
\end{figure}

\begin{figure}[!ht]
	\centering
	\includegraphics[width=0.45\textwidth]{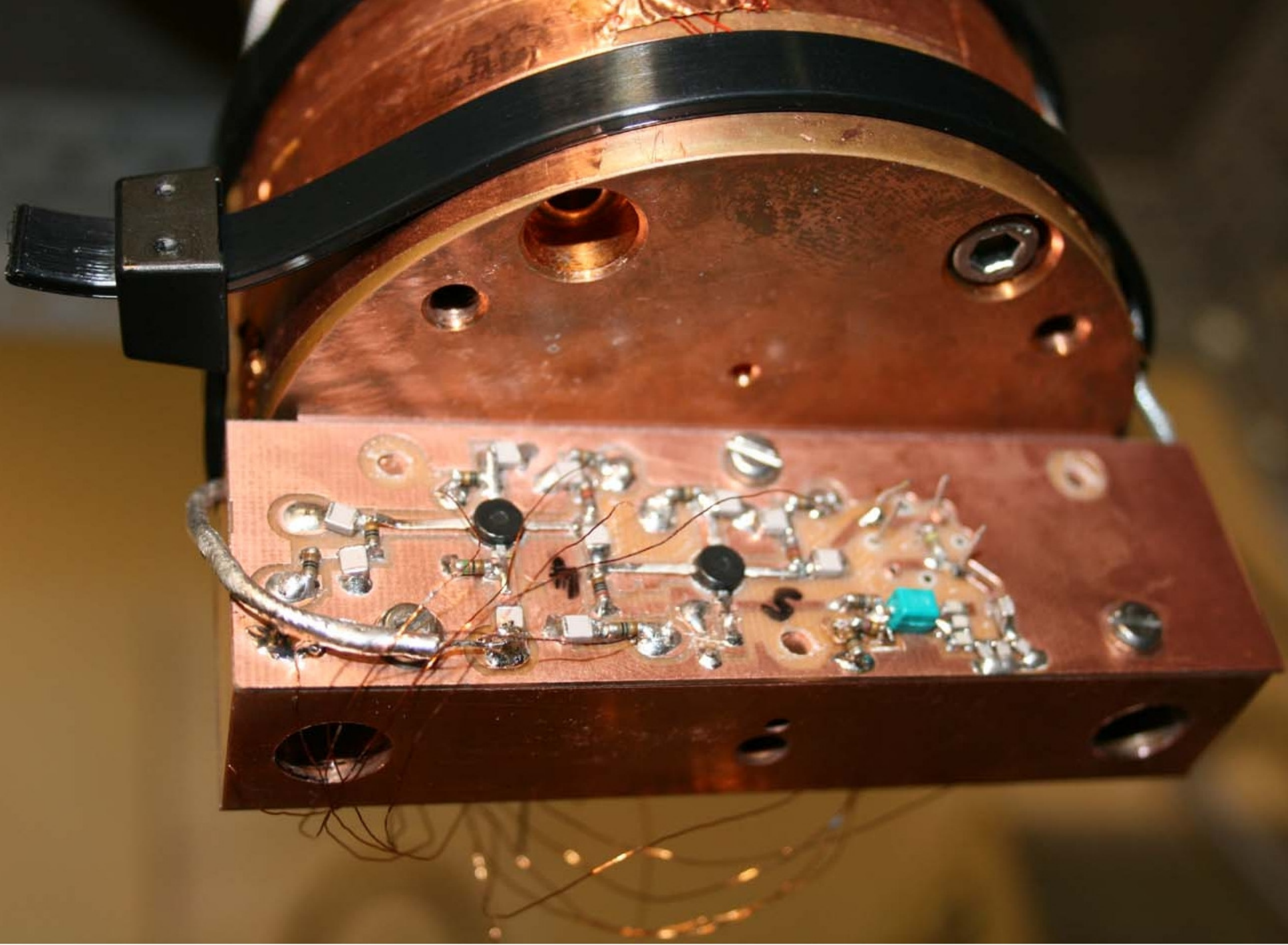}
	\caption{Oscillator mounting inside the cryocooler.}
\label{D002DO}
\end{figure}

The oscillator circuit board is mounted on the copper holder. Copper has a very high thermal conductivity ($1900$~$\frac{\text{W}}{\text{m}\cdot\text{K}}$) and a low specific heat ($0.1$ $\frac{\text{J}}{\text{kg}\cdot\text{K}}$) at $4$K, {which means that thermal resistance and capacitance between temperature control elements and the crystal resonator are minimized. The resonators are placed vertically in the special cavities in the copper holder that provide direct thermo-mechanical contact with the controlled media (see Fig.}~\ref{D001DO}{). This solution provides good temperature control of the crystal resonator. Vertical installation of resonators is needed to reduce cryocooler vibration noise. Electronic components other than the quartz resonator are separated from the holder by a layer of printed circuit board in order to reduce resonator heating. Environmental influences on the DUT are reduced by the anti-radiation shield which minimises the radiation heat exchange, and the vacuum chamber which acts like a dewar tube. Connecting cables and wires are attached to the copper block and the second and first stages in order to reduce thermal losses and fluctuations. A photograph of the installed oscillator is shown in Fig.}~\ref{D002DO}.

\begin{figure*}[!ht]
	\centering
	\includegraphics[width=0.65\textwidth]{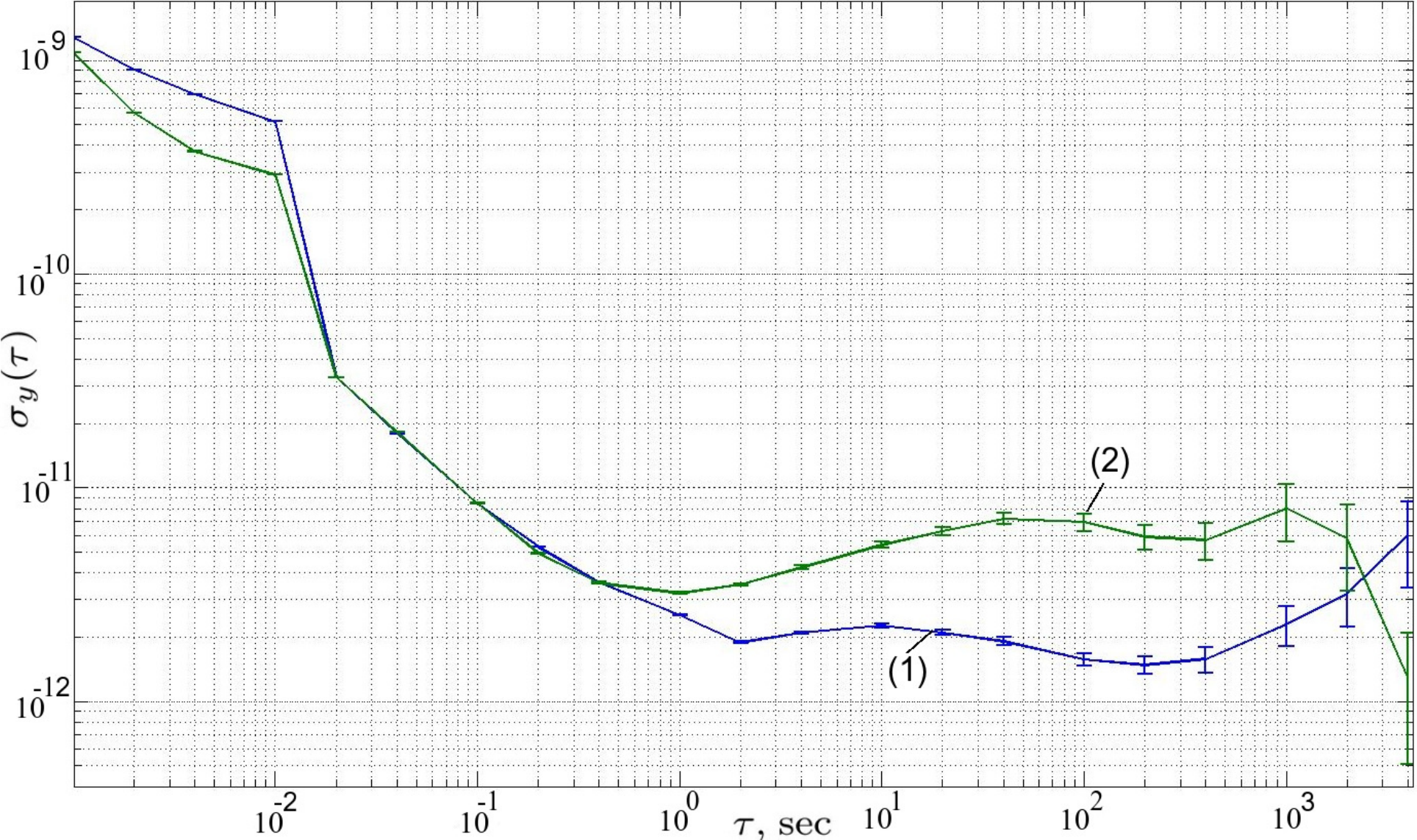}
	\caption{Allan deviation of two MOSFET-based liquid helium oscillators:  $8.4430359$~MHz (1) and $8.443156$~MHz (2).}
\label{D004DO}
\end{figure*}

\begin{figure*}[!ht]
	\centering
	\includegraphics[width=0.65\textwidth]{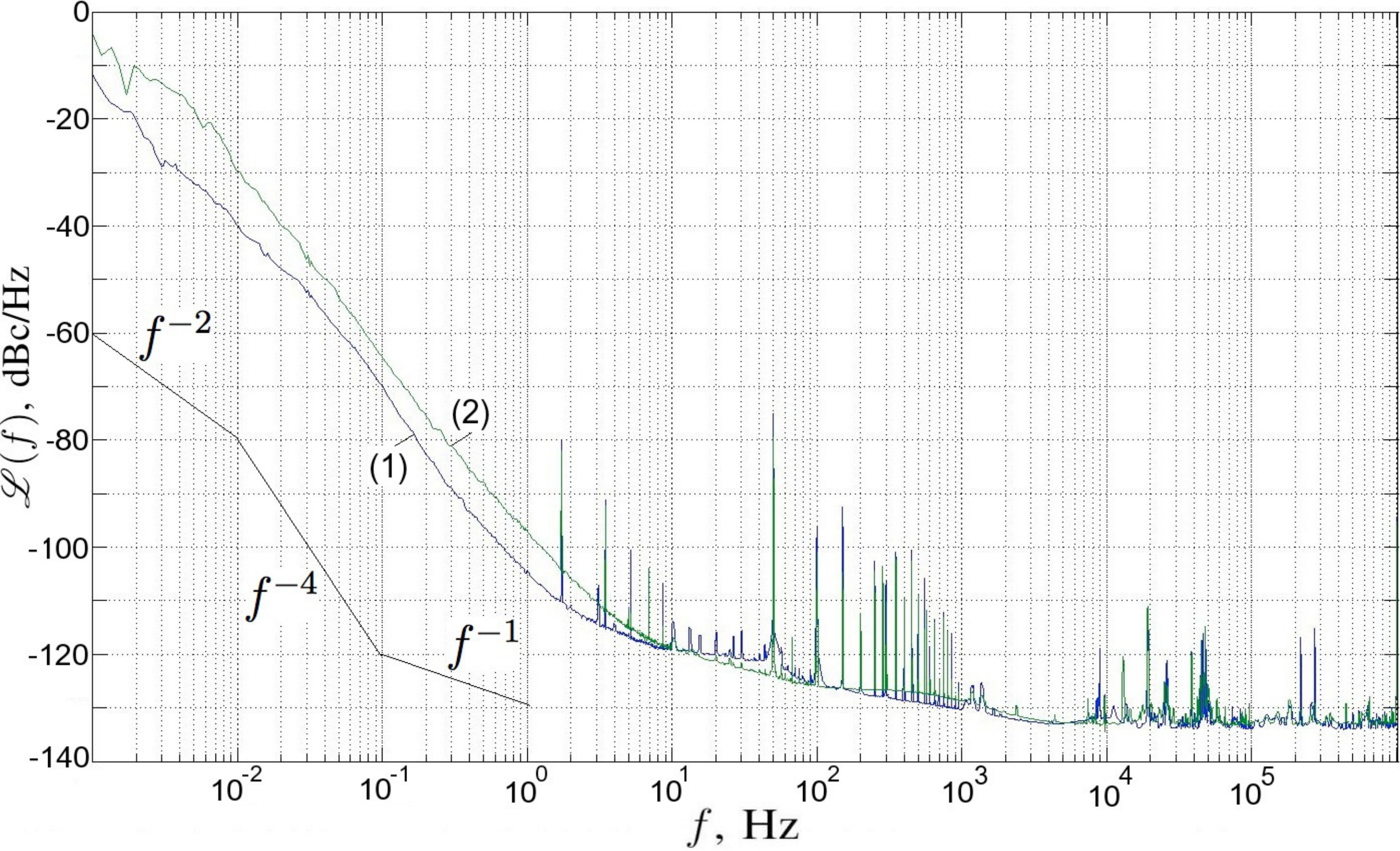}
	\caption{PSD of two MOSFET-based liquid helium oscillators:  $8.4430359$~MHz (1) and $8.443156$~MHz (2).}
\label{D005DO}
\end{figure*}

As predicted, several harmonics could be excited by changing the transistor bias voltages. These modes are $5.489680$~MHz (the 3$^{\text{rd}}$ overtone of the B mode), $4.992987$~MHz (the 3$^{\text{rd}}$ overtone of the C mode), $8.4430359$~MHz (an anharmonic), $8.535128$~MHz (an anharmonic), and $11.853208$~MHz (an anharmonic). All the attempts to excite the 5$^{\text{th}}$ overtone of the A mode (that with the highest $Q$ among 3$^{\text{rd}}$ and 5$^{\text{th}}$ overtones) were unsuccessful. The best results in terms of frequency stability were achieved with an anharmonic of $8.4430359$~MHz. The Allan deviation of two MOSFET-based oscillators is shown in Fig.~\ref{D004DO}. The corresponding phase noise power spectral density (PSD) is shown in Fig.~\ref{D005DO}. The results are obtained for these oscillators measured against an H-maser with the TSC5120A phase noise test set.

The phase noise PSD (Fig.~\ref{D005DO}) has an atypical order of spectrum slopes. Indeed, the random walk of phase is situated for lower Fourier frequencies $f^{-2}$ than the slope of $f^{-4}$. This suggest a violation of the Leeson principle for some oscillator instabilities. Besides that, a clear spurious frequency of $1.7$~Hz and its harmonics is seen. These spurious frequencies are the result of periodic cryocooler pulsations.

\section{Theoretical Explanation of the Measured Results}
\label{addii}

Fig. \ref{D005DO} shows that the closest to the carrier slope of the phase noise PSD is less (about $f^{-2}$) than the second one from the carrier ($f^{-4}$). These results may seem to be a violation of the well-known Leeson law~\cite{lees}. Nevertheless, they could be explained by revising the Leeson effect by the expansion of the theory on amplitude-phase relations in oscillators. This idea is briefly discussed in this section.

\begin{figure*}[t!]
        \centering
        \begin{subfigure}[b]{0.5\textwidth}
                \centering
                \includegraphics[width=0.7\textwidth]{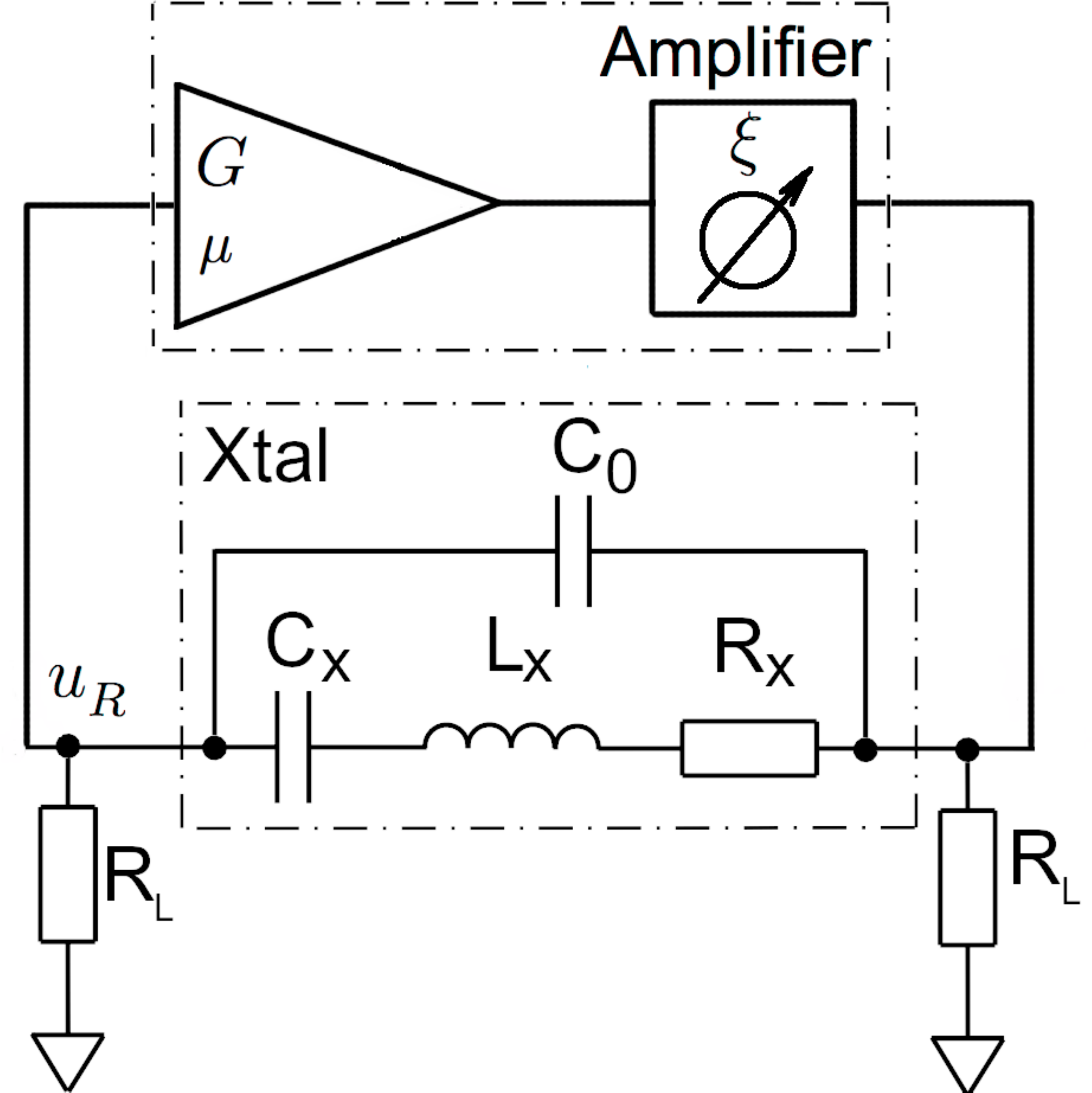}
               \caption{General oscillator model}
                \label{DF001HI:aa}
        \end{subfigure}%
        ~ 
        \begin{subfigure}[b]{0.5\textwidth}
                \centering
                \includegraphics[width=0.7\textwidth]{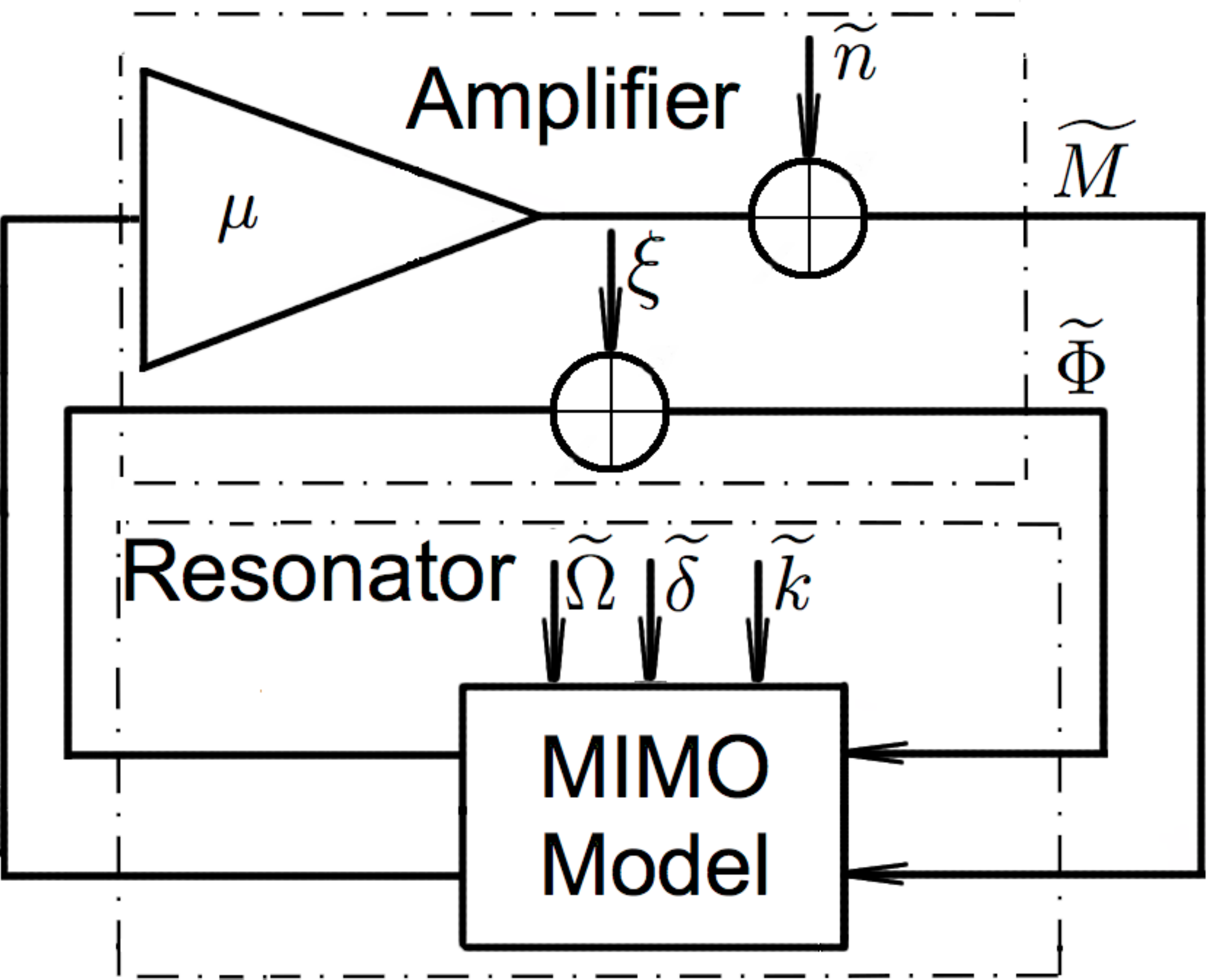}
                \caption{MIMO representation}
                \label{DF001HI:bb}
        \end{subfigure}
        \caption{Oscillator model for the phase noise analysis.}\label{DF001HI}
\end{figure*}

The analyzed oscillator model is shown in Fig.~\ref{DF001HI}. The system consists of a resonator inside a $\pi$-network and an amplifier. The resonator is represented with its linear MIMO model presented in~\cite{ultras}:
\begin{equation}
\mathbf{Y} = \mathbf{H}_i\mathbf{X}_i+\mathbf{H}_r\mathbf{X_r},
	\label{W200LP}
\end{equation}
where
\begin{equation}
	\label{DD002D}
\mathbf{Y}=\left( \begin{array}{l}
	\widetilde{M}_o\\
	\widetilde{\Phi}_o
 \end{array} \right),
 \mathbf{X}_i=\left( \begin{array}{l}
	\widetilde{M}\\
	\widetilde{\Phi}
	\end{array} \right),
 \mathbf{X}_r=\left( \begin{array}{l}
	\widetilde{\delta}\\
	\widetilde{\Omega}\\
	\widetilde{k}\\
\end{array} \right),
 \end{equation}
 are correspondingly a vector of small fluctuations of an output signal averaged over a period of the oscillation (amplitude and phase), fluctuations of a resonator input signal, and a vector of fluctuation of resonator internal parameters. Here $\widetilde{\delta}$, $\widetilde{\Omega}$ and $\widetilde{k}$. {Here and further, parameters are decomposed into constant $\overline{x}$ and small fluctuating $\widetilde{x}$ components.} are small fluctuations of the resonator damping, detuning frequency, and attenuation. $\mathbf{H}_i$ and $\mathbf{H}_r$ are transfer matrices for corresponding instabilities\cite{ultras, book1}.

In order to simplify the analysis, the amplifier is considered to have infinite input and output impedances, so the derived MIMO model is relevant. Moreover, the amplifier is modelled by a small signal gain $\mu$ and a phase shifter $\xi$. So, the amplifier transfer matrix in the domain of amplitude-phase fluctuations is
\begin{equation}
	\label{H200LP}
\mathbf{P} =\left( \begin{array}{ll}
	\mu & 0\\
	0 & 1
 \end{array} \right)
\end{equation}
The phase shifter is introduced in order to model the typical situation when oscillator does not work exactly at the resonance frequency. At the same time, the phase shifter does not increase the order of the model, so it does not effect its complexity. Moreover, since the amplifier is a wide-band component (compared to the resonator), the phase shifter is a good approximation of its dynamical properties. Decomposing $\xi$ into constant $\overline{\xi}$ and variable $\widetilde{\xi}$ parts, the latter part may be considered as the amplifier phase noise. Correspondingly, it is possible to introduce $\widetilde{n}$ as the oscillator amplitude noise. In addition to the small signal gain $\mu$, it is necessary to introduce a ``large signal" amplifier gain $G$, as a ratio between steady state amplitudes at the amplifier output and input. It can be shown that for typical oscillator nonlinearities, the ratio $\nu = \frac{\mu}{G}$ is always less than unity. 

{ The following analysis is based on perturbation approach, where large-signal steady-state parameters of oscillations ($G$, $\overline{\Omega}$, etc) are found on the first step, and small signal variables ($\widetilde{\Omega}, \widetilde{M}$, etc) are considered as small perturbations around steady-state solutions on the second stage. This approach linearises the problem allowing us to treat large-signal components simply as parameters.}

Fig.~\ref{DF001HI:bb} represents the oscillator in amplitude-phase space with the state vector $\mathbf{Y} = [\widetilde{M}\hspace{3pt} \widetilde{\Phi}]^T$. The system is disturbed by amplifier amplitude $\widetilde{n}$ and phase $\widetilde{\xi}$ noises as well as by resonator parameter fluctuations $\widetilde{\Omega}$, $\widetilde{\delta}$ and $\widetilde{k}$. Using this oscillator representation it is easy to find a transfer function from each noise source to the output signal phase $\widetilde{\Phi}$. These transfer functions are an extension of the standard Leeson effect on oscillator parameter fluctuation in the presence of the amplitude-phase coupling. They show the impact of the corresponding noise source into the oscillator closed loop phase noise. In fact, all these transfer functions could be classified in two kinds: 1) transfer functions that have an integrative denominator multiplier that introduces the change of PSD slope near the carrier (from $\widetilde{\xi}$, $\widetilde{\delta}$ and $\widetilde{\Omega}$); and 2) transfer functions that do not have an integrative multiplier. Thus, they do not change the PSD slope near the carrier (from $\widetilde{n}$, $\widetilde{k}$).

The resulting oscillator phase noise is a superposition of all phase noise spectra multiplied by the corresponding squared transfer functions. It is usually considered that the transfer functions of the first type dominate. However, it is possible that for some range of Fourier frequencies, noise sources of the second type of transfer function become dominant.

Let us consider the usual case for BAW oscillators where $f_c>f_L=\frac{1}{2\pi}\overline{\delta}$ (see Fig. 8: fL is for Leeson frequency and $f_c$ for corner frequency). To simplify the discussion, it is supposed without loss of generality that noise sources $\widetilde{\xi}$ and $\widetilde{k}$ (representing both transfer function types) are dominant over all the rest.
Their transfer functions are:
\begin{equation}
	\label{H201LP}
\left. \begin{array}{ll}
	\displaystyle \mathscr{H}_{\xi\Phi}(s) = \frac{\widetilde{\Phi}}{\widetilde{\xi}} = 1+\frac{\overline{\delta}}{s}+\frac{\overline{\Omega}^2}{s}\frac{1-\nu}{s+\overline{\delta}(1-\nu)},\\
	\displaystyle \mathscr{H}_{k\Phi}(s) = \frac{\widetilde{\Phi}}{\widetilde{k}} = -\frac{\overline{\Omega}}{\overline{k}}\frac{1}{s+\overline{\delta}(1-\nu)},
 \end{array} \right.
\end{equation}

Also, it is considered that $S_k(f)$ has a slope of $f^{-2}$ for close-to-carrier Fourier frequencies (this could be the case of random-walk due to temperature). In the same time $S_\xi(f)$ is proportional to $f^{-1}$ for frequencies less than the corner frequency $f_c$, and $f^0$ for higher Fourier frequencies.

\begin{figure*}[!ht]
	\centering
	\includegraphics[width=0.68\textwidth]{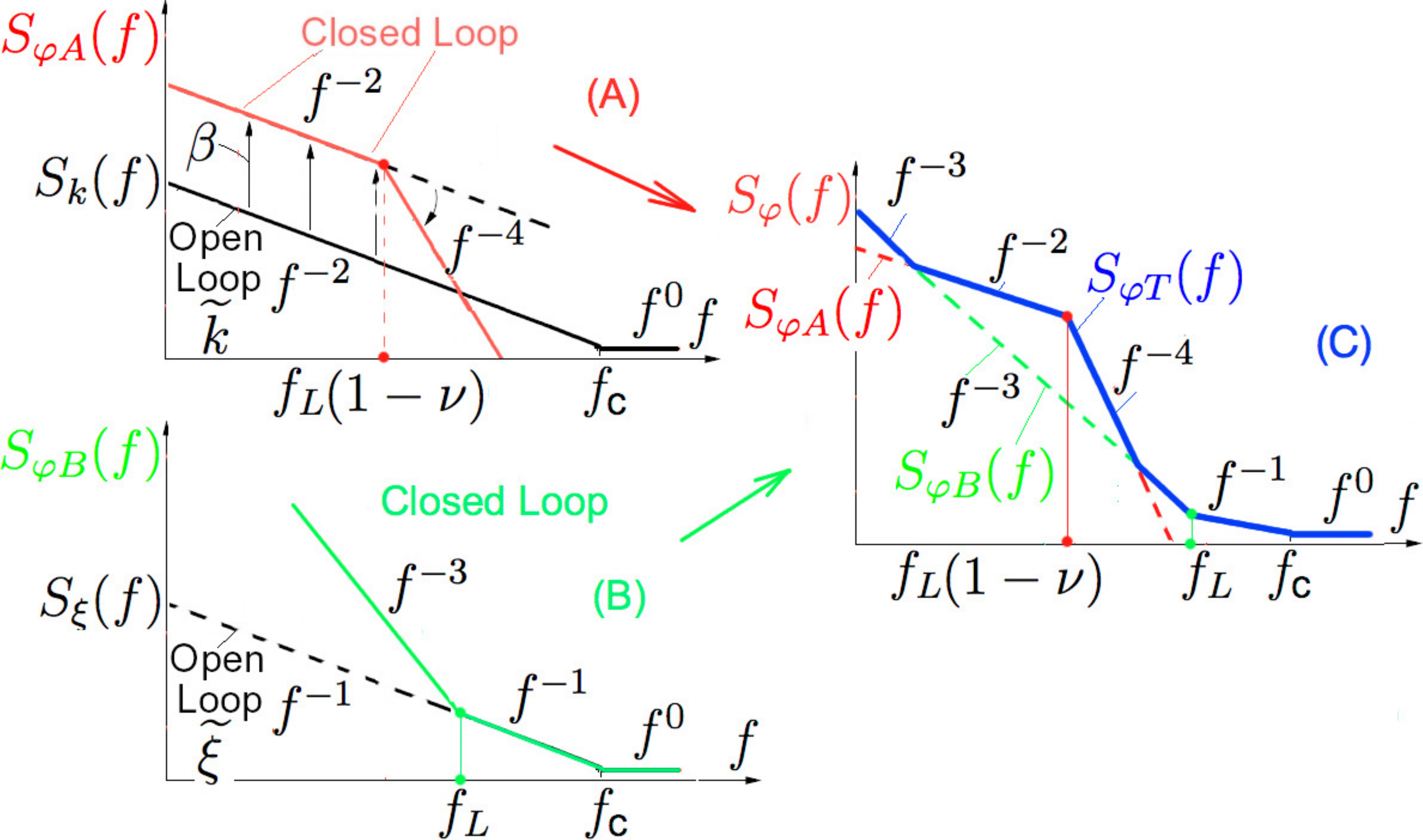}
	\caption{Predicted oscillator phase noise PSD for an oscillator with high $Q$: (A) effect of the closed loop (transfer function $\mathscr{H}_{k\Phi}$) on $S_k(f)$, (B) effect of the closed loop (transfer function $\mathscr{H}_{\xi\Phi}$) on $S_\xi(f)$, (C) - possible resulting PSD.}
\label{D00EDO}
\end{figure*}

The impact of the transfer function $\mathscr{H}_{k\Phi}$ on $S_k(f)$ is twofold. First, the transfer function multiplies the spectra by
\begin{equation}
\beta = \frac{\overline{\Omega}}{\overline{k}\overline{\delta}(1-\nu)}=\frac{2\overline{\Omega}}{\overline{k}\overline{\omega}_0(1-\nu)}\overline{Q}_L
\label{H202LP}
\end{equation}
inside the transfer function bandwidth. This means that for resonators with higher loaded quality factor $Q_L$ and the same resonance frequency $\overline{\omega}_0$ (this is the case for cryogenic oscillators), the corresponding noise multiplication coefficient is higher. Consequently, the noise impact into the total noise is more significant. Second, for frequencies higher than a cut of value $f_L (1-\nu)$, the spectrum is multiplied by $f^{-2}$ (outside the transfer function bandwidth). These transformations are shown in Fig. \ref{D00EDO}, (A). The transfer function $\mathscr{H}_{\xi\Phi}$ modifies the corresponding phase noise PSD according to the standard Leeson effect with a small correction term due to amplitude-phase conversion.

The resulting oscillator phase noise PSD is:
\begin{equation}
S_\varphi(f) = \mathscr{H}^2_{k\Phi}S_k(f)+\mathscr{H}^2_{\xi\Phi}S_\xi(f).
\label{H203LP}
\end{equation}
Let the resonator loaded quality factor be high enough. Then it is possible that the first term of equation~\ref{H203LP} overlaps the second. Thus, for some range of frequencies $S_A(f)$ hides the standard Leeson effect ($S_B(f)$) introducing two additional spectrum slopes ($f^{-2}$ and $f^{-4}$).

So, if $Q_L$ is high enough, then for some range of frequencies $f^{-2}$ and $f^{-4}$ noises of $S_A(f)$ dominate over $f^{-3}$. This situation gives the following order of slopes starting from the carrier:
\begin{enumerate}
\item $f^{-3}$ - the $\widetilde{\xi}$ flicker noise multiplied by $f^{-2}$ ($\mathscr{H}^2_{\xi\Phi}(f)f^{-1}$ for $f<f_L$);
\item $f^{-2}$ - the $\widetilde{k}$ random walk multiplied by $\beta$ ($\mathscr{H}^2_{k\Phi}(f)f^{-2}$ for $f<f_L(1-\nu)$);
\item $f^{-4}$ - the $\widetilde{k}$ random walk multiplied by $\beta f^{-2}$ ($\mathscr{H}^2_{k\Phi}(f)f^{-2}$ for $f>f_L(1-\nu)$);
\item $f^{-1}$ - the $\widetilde{\xi}$ flicker noise ($\mathscr{H}^2_{\xi\Phi}(f)f^{-1}$ for $f_L<f<f_c$);
\item $f^{0}$ - the $\widetilde{\xi}$ flicker noise ($\mathscr{H}^2_{\xi\Phi}(f)f^{-1}$ for $f>f_c$);

\end{enumerate}

So, the output phase noise PSD of an oscillator could be different from that predicted by the standard Leeson relations. This happens because standard formulation of the effect does not takes into account amplitude-phase relations and does not consider fluctuations of some important oscillator parameters. It is shown that even for feedback oscillators with usual topologies, a PSD could have additional spectrum slopes. This is the case of oscillators based on resonators with very high quality factors, such as the cryogenic oscillators presented in this paper. So, the cryogenic quartz oscillators exhibit effects unknown in room temperature devices.

\section{HBT-Based Design}
\label{HBT}

\begin{figure*}[!ht]
	\begin{center}
	\includegraphics[width=0.75\textwidth]{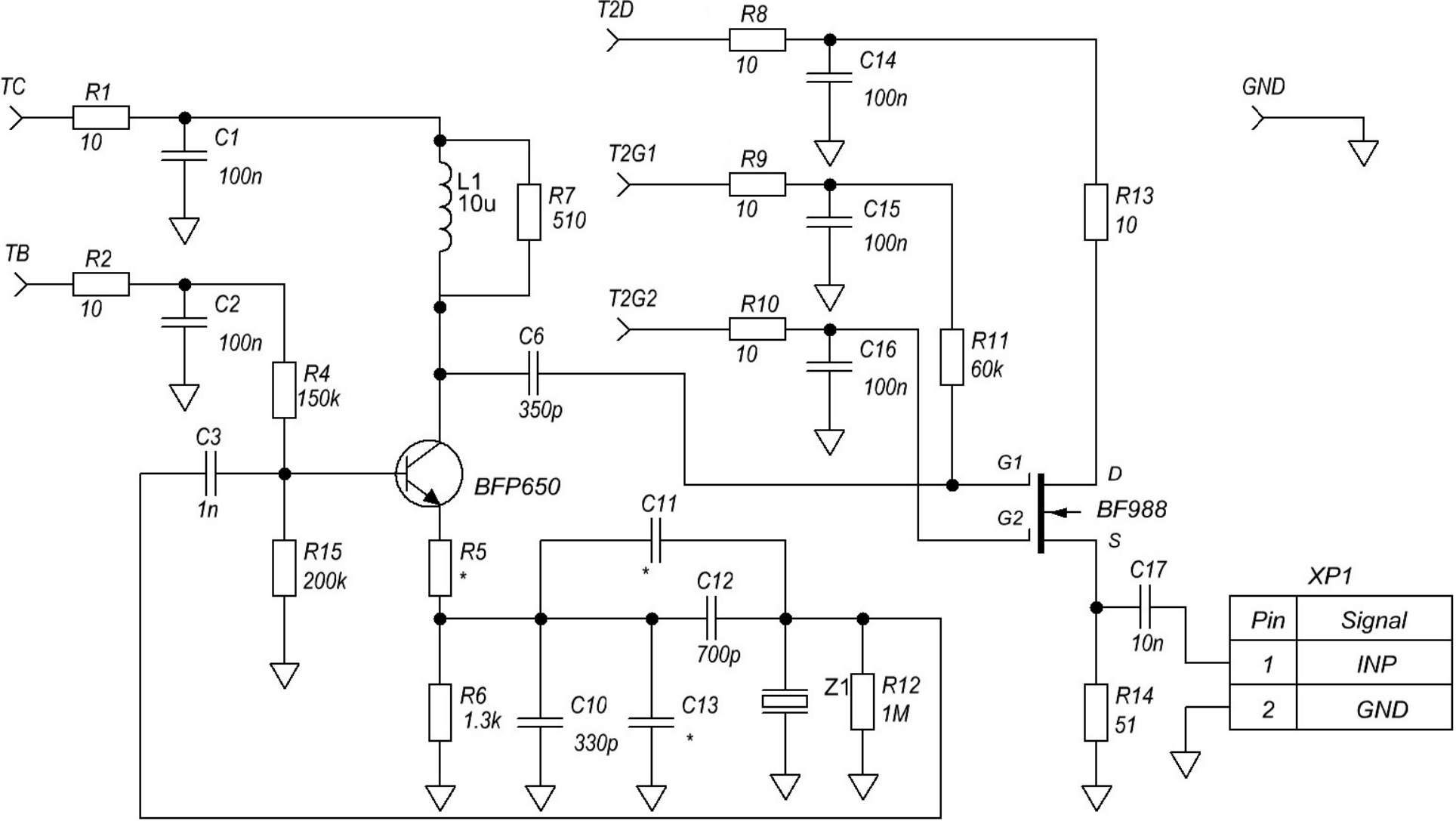}
	\caption{Schematic diagram of a cryogenic crystal oscillator based on a SiGe HBT.}
\label{D006DO}
	\end{center}
\end{figure*}

The same attempt of designing a liquid helium oscillator was undertaken using a bipolar transistor (in particular SiGe heterojunction bipolar transistor BFP650 from Infineon). Bipolar transistors are often preferable to field-effect devices in hyper stable $1-20$~MHz oscillators due { to better phase noise of the latter}. The same idea has been verified for liquid helium oscillators. The topology of such an oscillator is shown in Fig.~\ref{D006DO}. The circuit is a classical Colpitts oscillator with a single bipolar transistor in the loop. A separating buffer is implemented using the MOSFET device.

{ The packaging solution for this design is fully identical to that of the MOSFET-based oscillator (see Section}~\ref{MOSFET}).

\begin{figure*}[!ht]
	\begin{center}
	\includegraphics[width=0.55\textwidth]{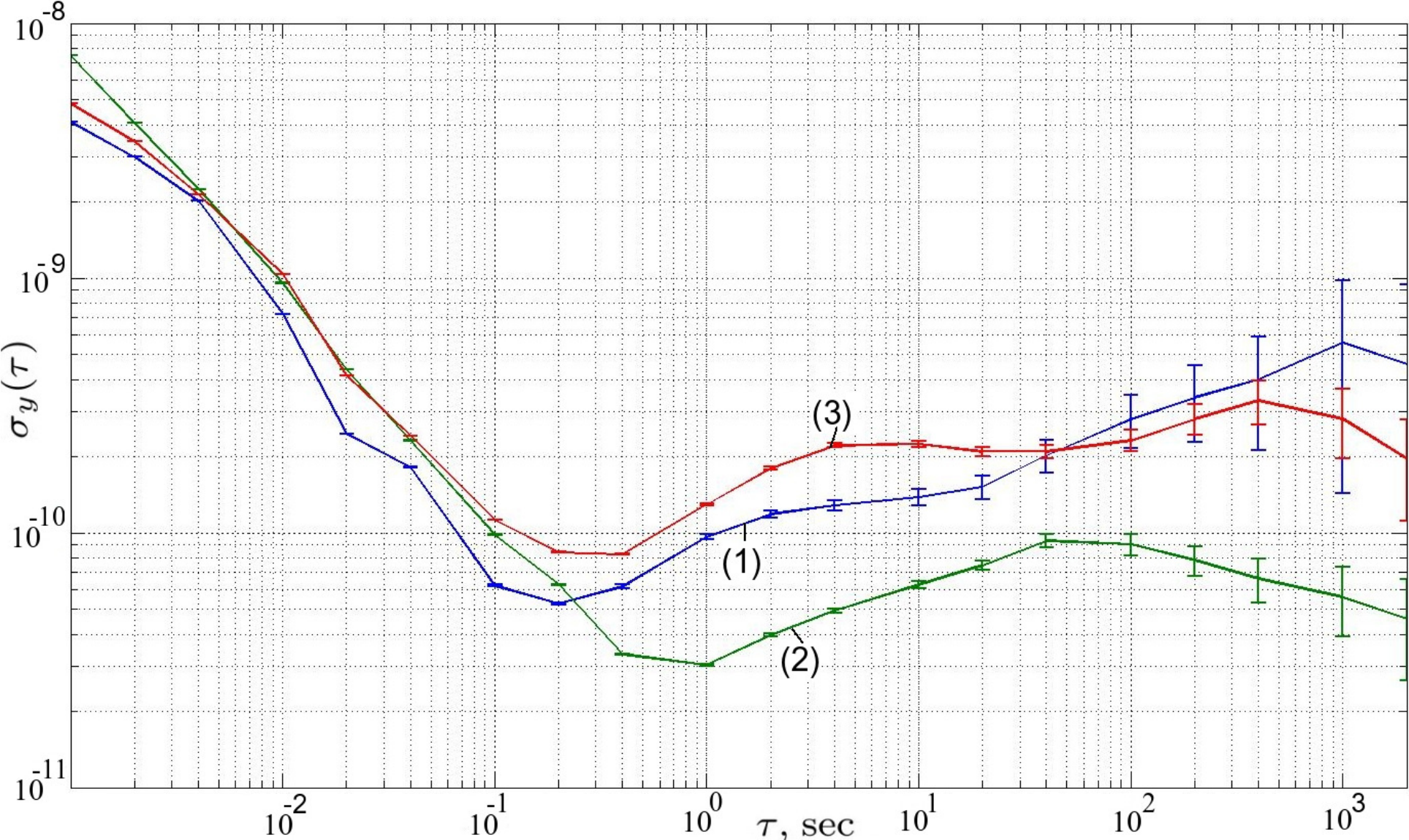}
	\caption{Allan deviation of two HBT-based liquid helium oscillators, for different bias voltages.}
\label{D007DO}
	\end{center}
\end{figure*}

\begin{figure*}[!ht]
	\begin{center}
	\includegraphics[width=0.55\textwidth]{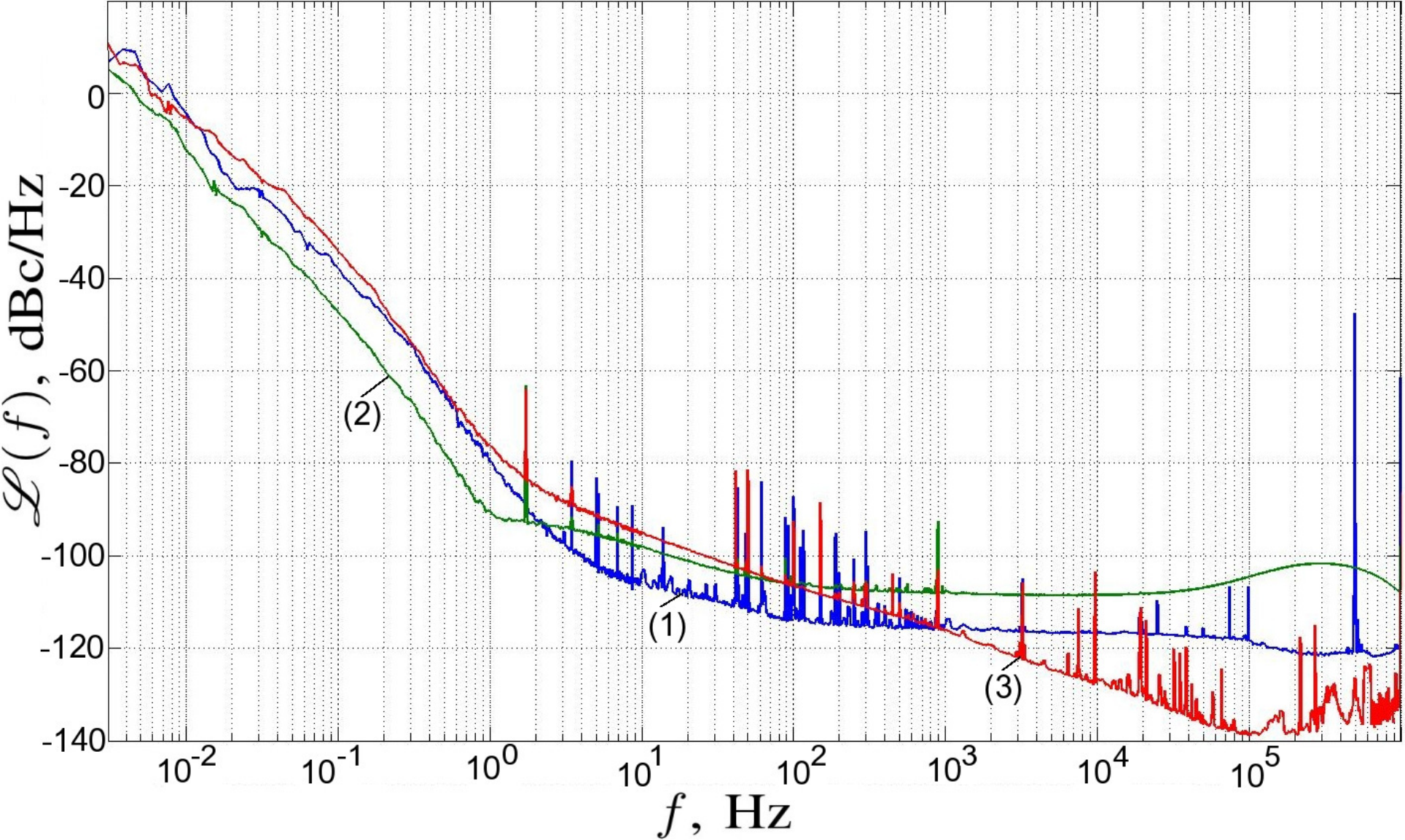}
	\caption{PSD of two HBT-based liquid helium oscillators, for different bias voltages.}
\label{D008DO}
	\end{center}
\end{figure*}

Like in the case of MOSFET oscillators, the HBT oscillator could be also excited on different frequencies depending on the bias voltages and regulation history. Among the possible oscillating frequencies are $5.489793$~MHz (the 3$^{\text{rd}}$ overtone of the B mode) and $4.992984$~MHz (the 3$^{\text{rd}}$ overtone of the C mode). The attempts to excite the 5$^{\text{th}}$ overtone of the A mode were also unsuccessful. The frequency stability measurement results in terms of the Allan deviation are shown in Fig.~\ref{D007DO}. The corresponding phase noise PSD is shown in Fig.~\ref{D008DO}.

It is clearly seen from Fig.~\ref{D008DO}  that the white phase noise is lower for the case of stronger oscillating signal (curve (3)) because of a better signal-to-noise ratio. At the same time, close-to-carrier instabilities are higher in such cases. Curve (2) is measured for the weakest signal. The same pattern is usually observed for room temperature oscillators.

Unfortunately, the oscillators tested are excited only on the 3$^{\text{rd}}$ overtone of the B mode and the 3$^{\text{rd}}$ overtone of the C mode of their resonators. This fact determines that their frequency stability results are inferior to that of the MOSFET-based cryogenic oscillators.
 
\section{Stability Limitations}
\label{SLDD}

As is evident from the experimental results obtained, the main problem of liquid helium oscillator development is the inability to excite the desired resonator modes. This problem due to two factors -- high dispersion of component characteristics and their high dependence on the exact environment and installation, as well as very limited means of development and optimization.

Nevertheless, the main problems of quartz resonators at liquid helium temperatures are their rather high self phase noise\cite{PhNoCr} and the absence of a turn-over point in frequency-temperature characteristics, as well as a strong amplitude-frequency effect\cite{SunFr}. These problems are serious obstacles on the way to developing hyper stable oscillators.

It is also worth noting that the cryogenic environment of the oscillator also has to be improved. Indeed, the cryocooler is a major source of external disturbances transferred to the oscillators through temperature fluctuations and vibrations.

\section{Conclusion}

The main conclusion of the work is that nowadays BAW and in particular quartz liquid helium oscillators are not able to compete with state-of-the-art atomic clocks and cryogenic sapphire oscillators. The main advantage of the cryogenic quartz devices -- very high quality factors -- is annulled by the absence of a frequency-temperature turnover point, and large nonlinear effects.

In addition, liquid helium temperature electronics remain an under-investigated field of science and engineering. Thus, the idea of fully cryogenic oscillators, as well as other systems, meets serious obstacles. There is a limited choice of electronic components, which exhibit unnatural behavior, high characteristic dispersion (for devices from the same series), and modelling approaches are currently unsophisticated. Moreover, the gap between simulation and experiment remains considerable. This leads to unpredictable results, such as, in the case of BAW oscillators, the inability to excite optimal modes.

{ Although at the stage of the first realisation, the BAW quartz technology cannot compete with the state-of-the-art cryogenic microwave sapphire oscillator\cite{CSO2008}.} Further development of cryogenic BAW devices must take into account the present results and aim to solve the stated problems. { The next generation of quartz frequency sources should be based on specially designed resonators with a frequency-temperature turnover point at $4$K, have passive and active vibration compensation system, employ more reliable active devices. }

\section*{References}
%

\end{document}